\documentstyle[aps,prl,epsf]{revtex}

\begin{document}
%\draft

%%% \twocolumn[{
%%% \widetext

\title{Green's Function Monte Carlo study of correlation functions in 
the (2+1)D $U(1)$ lattice gauge theory}

\author{C. J. Hamer\cite{email}, R. J. Bursill$^\dagger$ and 
M. Samaras}

\address{
School of Physics,
University of New South Wales,
Sydney, 2006, Australia.\\
$^\dagger$Department of Physics, UMIST, PO Box 88,
Manchester, M60 1QD, UK. 
}

\maketitle
\mediumtext

\begin{abstract}
A ``forward walking'' Quantum Monte Carlo (QMC) algorithm has been 
developed to calculate correlation functions for the Hamiltonian lattice 
formulation of $U(1)$ Yang-Mills theory in (2+1) dimensions. It is shown 
that Wilson loops can be calculated with high accuracy. Creutz ratios 
are used to determine the string tension, which agrees with results from 
other approaches. Timelike correlations are used to estimate the mass 
gaps, which agree with series expansion results in the strong coupling 
regime.
\end{abstract}

%%% }]
%%% \narrowtext

%\widetext

\section{Introduction}
\label{sec:intro}

The two major variants of lattice gauge theory (LGT) are the 
``Euclidean'' formulation of Wilson \cite{wilson}, and the 
``Hamiltonian'' version of Kogut and Susskind \cite{kogut}. In the 
Euclidean r\'{e}gime, classical Monte Carlo simulations have proved to 
be extremely powerful in extracting quantitative predictions from the
theory, as first shown by Creutz \cite{creutz}. This approach is
preferred by an overwhelming majority of lattice gauge theorists at the
present time.

The Hamiltonian formulation is still worthy of study, however. It can 
provide a valuable check of universality, for instance. Lattice gauge 
theory relies on the fundamental assumption that quantities such as mass 
ratios calculated in the continuum limit (a critical point of the 
lattice model) must be `universal', i.e.\ independent of the microscopic 
lattice structure or space-time formulation. There is not much real 
doubt that this is correct, but it is important to provide checks where 
possible \cite{hamer0}. Another reason is that many techniques imported 
from quantum many-body theory and condensed matter physics can be 
employed in this arena, which may give useful results. Examples include 
the strong-coupling series approach \cite{banks}, the $t$-expansion 
\cite{horn}, the coupled-cluster method \cite{guo,bishop}, and others. 
Nevertheless, it seems likely that Monte Carlo simulations will provide 
the most robust and accurate numerical techniques in this area also. Our 
aim in this paper is to discuss some further applications of these 
Quantum Monte Carlo methods \cite{kalos}.

The use of quantum Monte Carlo methods in Hamiltonian LGT has a
long and somewhat chequered history, and lags a good ten years behind
the Euclidean developments. The first calculations used a
strong-coupling basis involving discrete ``electric field'' link
variables, and a ``Projector Monte Carlo'' approach 
\cite{blank,degrand}, which used the Hamiltonian itself to project out 
the ground state. A later version of this was the ``stochastic 
truncation'' approach of Allton et al.\ \cite{allton}. Using this 
approach one can successfully compute string tensions and mass gaps for 
Abelian models \cite{hamer1}. For non-Abelian models, however, some
problems arose 
\cite{hamer1}. Using an electric field representation for the link 
variables and a Robson-Webber recoupling scheme \cite{robson} at the 
vertices requires the use of Clebsch-Gordan coefficients or 
6$j$-symbols, which are not known to high order for $SU(3)$; and 
furthermore, the `minus sign' problem rears its head, in that 
destructive interference occurs between different paths to the same 
final state. It may well be that a better choice of strong-coupling 
basis, such as the `loop representation', might avoid these problems; 
but this has not yet been demonstrated.

Heys and Stump \cite{heys1} and Chin et al.\ \cite{chin1} pioneered the 
use of ``Greens Function Monte Carlo'' (GFMC) or ``Diffusion Monte
Carlo'' techniques in Hamiltonian LGT, in conjunction with a
weak-coupling representation involving continuous gauge field link
variables. This was successfully adapted to non-Abelian Yang-Mills
theories \cite{heyschin2}, with no minus sign problem arising. In this
representation, however, one is simulating the wave function in gauge
field configuration space by a discrete ensemble or density of random
walkers: it is not possible to determine the derivatives of the gauge
fields for each configuration, or to enforce Gauss's law explicitly, and
the ensemble always relaxes back to the ground state sector. Hence one
cannot compute the string tensions and mass gaps directly as Hamiltonian
eigenvalues corresponding to ground states in different sectors, as one
does in the strong-coupling representation. Chin, Long and Robson
\cite{heyschin3} thus resorted to ``variational Monte Carlo'' (VMC)
techniques to compute mass gaps. They obtained some reasonable results;
but this approach always suffers from the major drawback that there is
an unknown systematic error in the results, due to their dependence on
the form of the variational wave function which is chosen.

It appears, therefore, that to make unbiased measurements of mass gaps 
in the weak-coupling representation one is forced back to the more 
laborious approach used in Euclidean calculations: namely, to measure an 
appropriate correlation function, and estimate the mass gap as the 
inverse of the correlation length. In ref.\ \cite{hamer1} the GFMC
method was tested on the (2+1)D U(1) model using a ``secondary 
amplitude'' technique to compute expectation values: but this 
proves to be expensive and prone to bias \cite{hamer3,samaras}.
 In this paper we will show how the standard `forward-walking' technique 
used in many-body theory \cite{kalos1} can be used for this purpose. The 
forward-walking method has already been applied to lattice spin models 
by Runge \cite{runge} and Samaras and Hamer \cite{samaras}. Here we 
apply it to the compact $U(1)$ Yang-Mills theory in (2+1) dimensions, 
which has been a standard test-bed for Hamiltonian LGT.

A brief discussion of the $U(1)$ model is given in Section II. Our Monte 
Carlo methods are outlined in Section III. The GFMC method is briefly 
summarized, and then the forward-walking method for estimating 
expectation values is discussed, together with a technique for measuring 
timelike correlations. In Section IV the results are presented. Our 
conclusions are summarized in Section V.

\section{Model Hamiltonian}
\label{sec2}

\begin{figure}[htbp]
%%%
 \centerline{\epsfxsize=8.4cm \epsfbox{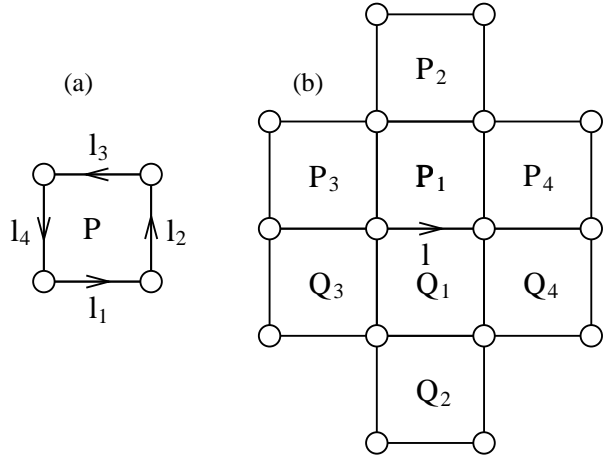}}
\caption{
(a) The links $l_1$,\ldots,$l_4$ associated with a plaquette $P$ on the 
lattice. (b) The plaquettes $P_1(l)$,\ldots,$P_4(l)$ and 
$Q_1(l)$,\ldots,$Q_4(l)$ associated with a link $l$ on the lattice.
}
\label{fig1}
\end{figure}

In a weak coupling basis the Hamiltonian for the compact $U(1)$ LGT in 
(2+1)D is given by \cite{kogut,drell}:
\begin{equation}
 H = -\sum_l \frac{\partial^2}{\partial A_l^2}
- 2x \sum_P \cos \theta_P
\label{eq1}
\end{equation}
where $A_l$ is the gauge field variable on link $l$ and
\begin{equation}
\label{eq2}
\theta_{P} \equiv A_{l_1} + A_{l_2} - A_{l_3} - A_{l_4} 
\end{equation}
is the plaquette variable for a lattice plaquette $P$, formed by the
four links $ l_1,\ldots,l_4 $, as illustrated in Fig.\ \ref{fig1}(a). We 
consider a periodic, square lattice of linear dimension $L$ and lattice 
spacing $a$. The `strong coupling' parameter $x = 1/e^4 a^2 = 1/g^{4}$ 
approaches infinity in the continuum limit $ a \rightarrow 0$.

This is an interesting model, which possesses some important 
similarities with QCD (for a more extensive review, see for example 
ref.\ \cite{hamer1}). If one takes the `naive' continuum limit at a 
fixed energy scale, one regains the simple continuum theory of
non-interacting photons \cite{gross}; but if one renormalizes or 
rescales in the standard way so as to maintain the mass gap constant,
then one obtains a confining theory of free massive bosons, as discussed
by Polyakov \cite{polyakov}, and proven by G\"{o}pfert and Mack 
\cite{gopfert_and_mack}. The Hamiltonian version of the model has been 
well studied by a variety of methods: some of the more recent include 
series expansions \cite{series,hamer2}, finite-lattice techniques 
\cite{irving}, the $t$-expansion \cite{horn1,morningstar}, and 
coupled-cluster techniques \cite{dabringhaus,fang,baker}, as well as QMC 
\cite{chin1,koonin,yung,hamer1}. Quite accurate estimates have been 
obtained for the string tension and mass gaps, which can be used as 
comparisons for our Monte Carlo results. The finite-size scaling 
properties of the model can be predicted using an effective Lagrangian 
approach combined with a weak-coupling expansion \cite{hamer4}, and the 
predictions agree very well with finite-lattice data \cite{hamer1}. 

\section{Monte Carlo Methods}
\label{sec3}

\subsection{Greens Function Monte Carlo}
\label{subsec: GFMC}

We use the Green's Function Monte Carlo [GFMC] method \cite{kalos}, 
which was adapted to the $U(1)$ model by Heys and Stump \cite{heys1}, 
and Chin et al.\ \cite{chin1}. A brief summary of the method can be 
given as follows.

In a weak-coupling representation, the basis states are taken to be
eigenstates of the plaquette angles $\theta_{P}$, which can take
continuous values. The Hamiltonian (\ref{eq1}) can be written as
\begin{equation}
H = - \sum_{l}\frac{\partial^{2}}{\partial A_{l}^{2}} + V(\Theta) \ ,
\label{eq3}
\end{equation}
where
\begin{equation}
V(\Theta) = -2 x \sum_{P} \cos \theta_{P},
\label{eq4}
\end{equation}
and the plaquette angles $\theta_{P}$ and link angles $A_{l}$ are 
related by equation (\ref{eq2}). The imaginary time Schr{\" o}dinger 
equation for the system is
\begin{equation}
-\frac{\partial}{\partial \tau} \Phi (\Theta ,\tau) =
[-\sum_{l}\frac{\partial^{2}}{\partial A_{l}^{2}} + V(\Theta ) - 
E_{\text{T}}]
\Phi (\Theta ,\tau), 
\label{eq6}
\end{equation}
where $E_{\text{T}}$ is a trial energy, representing a constant shift in 
the zero of energy, which will prove useful. The imaginary time 
evolution operator $\exp[-(H-E_{\text{T}})\tau]$ acts as a projector 
onto the ground state $ |\Phi_{0} \rangle $:
\begin{equation}
|\Phi_{0} \rangle = \lim_{\tau \rightarrow \infty}
e^{- \tau (H - E_{\text{T}} } \left| \Psi_{\text{T}} \right\rangle 
\label{eq6a}
\end{equation}
for any trial state $|\Psi_{\text{T}} \rangle$, provided that 
$|\Psi_{\text{T}} \rangle$ is not orthogonal to $|\Phi_{0}\rangle$.

Equation (\ref{eq6}) is a diffusion equation in configuration space, and 
is easily simulated by the Green's Function Monte Carlo method. It is
assumed that the ground-state wave function can be chosen positive
everywhere, and it is simulated by the density distribution of an
ensemble of random walkers $\{\Theta_{i}\}$ in configuration space,
with weights \{$ w_{i}$\}. The first term on the right of Eq.\ 
(\ref{eq6}) produces diffusion, and is simulated by a Gaussian random 
walk of the members of the ensemble as time proceeds, while the term 
$ [V(\Theta ) - E_{\text{T}} ] $ produces a growth or decay in the 
density which is simulated by a branching process.

\subsection{Variational Guidance}
\label{subsec: var}

The efficiency and accuracy of the simulation are greatly enhanced by
the use of variational guidance or importance sampling \cite{kalos}. Let 
$\Psi_{\text{T}} (\Theta)$ be a variational approximation to the true
ground-state wave function, and define a new probability distribution 
\begin{equation}
f(\Theta,\tau) = \Phi (\Theta,\tau) \Psi_{\text{T}}(\Theta), 
\label{eq7}
\end{equation}
Then the modified imaginary time Schr{\"o}dinger equation for
$f(\Theta,\tau)$ reads
\begin{equation}
-\frac{\partial f}{\partial \tau} =
-\sum_{l}\frac{\partial^{2}f}{\partial A_{l}^{2}}
+ [ E_{\text{L}}(\Theta) - E_{\text{T}} ] f +
\sum_{l} \frac{\partial}{\partial A_l} (f F_{Ql}(\Theta)), 
\label{eq8}
\end{equation}
where
\begin{equation}
E_{\text{L}} (\Theta ) =
\frac{ 1 }{ \Psi_{\text{T}}(\Theta) } H \Psi_{\text{T}}(\Theta)
\label{eq9}
\end{equation}
is the local energy obtained from the trial function, and
\begin{equation}
F_{Ql} (\Theta ) \equiv
\frac{ 2 }{ \Psi_{\text{T}}(\Theta) }
\frac{\partial \Psi_{\text{T}}(\Theta)}{\partial A_{l}} 
\label{eq10}
\end{equation}
is a ``quantum force'' term, which produces a directed drift in the 
ensemble towards the configurations favoured by the trial wave function. 
By a good choice of $\Psi_{\text{T}} $ and $ E_{\text{T}} $ the ``excess 
local energy'' term $[ E_{\text{L}}(\Theta) - E_{\text{T}} ]$ can be 
made very small, which reduces the amount of branching necessary, and 
reduces the statistical fluctuations in the results.

For small time steps $\Delta \tau$, an approximate Green's function
solution to Eq.\ (\ref{eq8}) is
\begin{eqnarray}
G(\Theta - \Theta', \Delta \tau) & \simeq & 
          \exp \{ -[ E_{\text{L}}(\Theta) - E_{\text{T}} ] \Delta \tau 
\} \\
\nonumber
& & \times \sum_{l}
\left(
      \frac{ 1 }{ {\surd \overline{4 \pi \Delta \tau} } }
      \exp \left\{
 -[ A_{l}' - A_{l} - \Delta \tau F_{Ql}(\Theta ) ]^{2} / 4 \Delta \tau
          \right\}
\right).
\label{eq11}
\end{eqnarray}
In the Monte Carlo simulation, each iteration corresponds to a time step
$\Delta \tau$. At each iteration, we sweep through each link in turn, 
and simulate the corresponding exponential factor in the sum on the 
right of (\ref{eq11}) by a random displacement of the link variable for 
each walker:
\begin{equation}
\Delta A_{l} = \Delta \tau F_{Ql}(\Theta) + \chi,
\label{eq12}
\end{equation}
where $\chi$ is randomly chosen from a Gaussian distribution with
standard deviation $\surd{\overline{(2 \Delta \tau)}}$. The first term 
in (\ref{eq12}) is the ``drift'' term, and the second is the 
``diffusion'' term. The first exponential on the right of (\ref{eq11}) 
is simulated by simply multiplying the ``weight'' of each walker $w_i$ 
by an equivalent amount.

At the end of each iteration, the trial energy $E_{\text{T}} $ is 
adjusted to compensate for any change in the total weight of all walkers 
in the ensemble; and a ``branching'' process is carried out, so that 
walkers with weight greater than (say) 2 are split into two new walkers, 
while any two walkers with weight less than (say) 1/2 are combined into 
one, chosen randomly according to weight from the originals. This 
procedure of ``Runge smoothing'' \cite{runge} maximizes statistical 
accuracy by keeping the weights of all the walkers within fixed bounds, 
while mimimizing any fluctuations in the total weight due to the 
branching process.

When equilibrium is reached after many sweeps through the lattice, the
average value of the trial energy $E_{\text{T}}$ will give an estimate 
of the ground-state energy $E_0$, and the weight density of the ensemble 
in configuration space will be proportional to $\Phi_0\Psi_{\text{T}}$. 
Various corrections due to the finite time interval $\Delta \tau$ have
been ignored in this discussion, and the limit
$\Delta \tau \rightarrow \infty$ must be taken in some fashion to 
eliminate such corrections.

In the simulations presented here, a trial function for the ground state 
was chosen as
\begin{equation}
\psi_{\text{T}}(\Theta) =
\exp \left[
c \sum_P \cos \theta_P + d \sum_{<PP'>} \cos( \theta_P + \theta_{P'} )
\right]
\label{eq13}
\end{equation}
with two variational parameters $c$ and $d$, where the sum over 
$\langle P P' \rangle$ denotes a sum over nearest-neighbour pairs of 
plaquettes, forming rectangles ($1 \times 2$ Wilson loops) on the 
lattice. That is, the uncorrelated single plaquette trial function used 
in \cite{hamer1} has been extended to include a {\em correlated} term. 
In the single plaquette case ($d = 0$) it is straightforward to 
determine the optimal value for $c$ \cite{heys1}. For the correlated 
case ($d \neq 0$), the optimum values for $c$ and $d$ must be found by 
means of a separate Variational Monte Carlo (VMC) calculation, where the 
variational energy
\begin{equation}
E_{\text{VMC}}(c,d) =
\frac{
\int ( \Psi_{\text{T}}(\Theta) )^2 E_{\text{L}}(\Theta) \; D\Theta
}
{
\int ( \Psi_{\text{T}}(\Theta) )^2 \; D\Theta
}
\end{equation}
is minimised. The procedure is described in \cite{heys1} where a six-
parameter variational wave function was used to estimate the mass gaps 
in the $U(1)$ model at the VMC level. In short, one samples the local 
trial energy over configurations $\{ \Theta \}$ randomly generated by a 
Metropolis procedure, distributed with respect to the weight function
$( \Psi_{\text{T}}(\Theta) )^2$. A downhill simplex method was used to 
perform the minimisation. To this end $E_{\text{VMC}}(c,d)$ was 
evaluated in a region centered around a guessed minimum $(c_0,d_0)$ 
using the reweighting procedure described in \cite{heys1}, whereby 
uncorrelated configurations generated with respect to the weight 
function $( \Psi( \Theta )_{c = c_0, d = d_0} )^2$ were reweighted to 
give a distribution corresponding to the required weight function for a 
pair $(c,d)$ in the neighbourhood of $(c_0,d_0)$. This procedure was 
iterated until the minimum $(c_0,d_0)$ converged. We used 100,000 
independent configurations on an $8 \times 8$ lattice to determine 
$E_{\text{VMC}}(c,d)$. This allowed the optimal $c$ and $d$ to be fixed 
to within around 1\%. Though it is possible to optimise $c$ and $d$ for 
each lattice size, for a given value of the coupling $x$, we used the 
same values of $c$ and $d$ for all lattice sizes $L$. The values of $c$ 
and $d$ used are listed in Table \ref{table0}.

The use of a correlated trial wave function is expected to markedly 
decrease the variance of estimators and the amount of branching in the 
population smoothing process, particularly in the physically interesting 
weak coupling regime ($x$ large). Moreover, because the trial function 
is closer to the true ground state, fewer forward walking iterations 
(see section \ref{subsec: for}) are required in order to derive 
converged estimators for correlation functions. These advantages come at 
a price in that updating of GFMC configurations becomes more 
complicated: The local trial energy is
\begin{equation}
E_{\text{L}}(\Theta)  =  ( 4 c - 2 x ) \sum_P \cos \theta_P
                      + 6 d \sum_{\langle P P' \rangle}
                              \cos \left( \theta_P + \theta_{P'} \right)
                    - \frac{1}{4} \sum_l F_{Ql}^2,
\end{equation}
and the quantum force term is given by
\begin{eqnarray}
F_{Ql} &  = & 
            c \left( \sin \theta_{P_1(l)} - \sin \theta_{Q_1(l)} \right)
       +\; d \sum_{j = 2}^4 \left\{
            \sin \left( \theta_{P_1(l)} + \theta_{P_j(l)} \right) 
\right.
\nonumber
\\ & & 
       -\; \left. \sin \left( \theta_{Q_1(l)} + \theta_{Q_j(l)} \right)
                           \right\}.
\end{eqnarray}
Here, $P_1(l)$,\ldots,$P_4(l)$ and $Q_1(l)$,\ldots,$Q_4(l)$ denote the 4
closest plaquettes on either side of the link $l$, as illustrated in 
Fig.\ \ref{fig1}(b). These expressions are more difficult to compute 
than in the single parameter ($d = 0$) case. However, the increased 
complexity is easily offset by the gains from variance reduction.

\subsection{Forward-walking method}
\label{subsec: for}

The quickest method of estimating an expectation value 
$\langle Q \rangle_{0}$ is simply to form the weighted average of $Q$ 
over the ensemble of random walkers (we assume Q is diagonal in the 
chosen basis, for simplicity). This produces an estimate according to 
the distribution $\Psi_{\text{T}}\Phi_{0}$, rather than $\Phi_{0}^{2}$. 
The estimate can be perturbatively improved \cite{kalos}, but there 
remains an unknown systematic error due to the dependence on the trial 
function $\Psi_{\text{T}}$. For this reason, we have preferred to use 
the so-called ``forward-walking'' method for estimating expectation 
values.

The forward-walking method is a robust technique for estimating 
expectation values \cite{kalos1,liu,whitlock}, based on the following 
equation \cite{runge} for an operator $Q$:
\begin{eqnarray}
<Q>_{0} & = & \frac{ \langle \Phi_{0} |Q| \Phi_{0} \rangle }
                   { \langle \Phi_{0}| \Phi_{0} \rangle } 
\label{eq14}
\\
        &
\stackrel{\textstyle \sim}{\scriptscriptstyle J \rightarrow \infty}
           &
\frac{ \langle \Psi_{\text{T}} | K^{J} Q | \Phi_{0} \rangle }
{ \rangle \Psi_{\text{T}} | K^{J} | \Phi_{0} \rangle }
\label{eq15}
\\
       & = &
\frac{ \sum \tilde{K}(\Theta_{J}, \Theta_{J-1}) \ldots 
\tilde{K}(\Theta_{2}, \Theta_{1})
Q(\Theta_{1}) \tilde{\Phi}_{0}(\Theta_{1})
     }{
\sum \tilde{K}(\Theta_{J}, \Theta_{J-1}) \ldots
\tilde{K}(\Theta_{2}, \Theta_{1}) \tilde{\Phi}_{0}(\Theta_{1})
      }
\label{eq16}
\end{eqnarray}
where $K(\Theta_{J},\Theta_{J-1})$ is the evolution operator for time 
$\Delta \tau$, and $\tilde{K}(\Theta_{J},\Theta_{J-1})$ is the same 
operator in the similarity transformed basis. Again we have assumed that 
the operator $Q$ is diagonal in the basis of plaquette variables 
$\Theta$.

This equation is implemented by \cite{kalos1,liu,whitlock} the following 
procedure:
\begin{itemize}
\item[i)] Starting from the trial state, iterate until equilibrium is
achieved, then begin a measurement;
\item[ii)] Record the value $Q(\Theta_{i})$ for each walker 
(``ancestor'') at the beginning of the measurement;
\item[iii)]
Propagate the ensemble as normal for $J$ iterations, keeping a record of 
the ``ancestor'' of each walker in the current population;
\item[iv)]
Take the weighted average of the $Q(\Theta_{i})$ with respect to the 
weights of the descendants of $\Theta_{i}$ after the $J$ iterations, 
using sufficient iterations $J$ that the estimate reaches a `plateau'.
\end{itemize}

This procedure has been tested for the magnetization in the 2D lattice 
Heisenberg model by Runge \cite{runge}, and for correlation functions 
in the 1D transverse Ising model by Samaras and Hamer \cite{samaras}, 
and works very well. The drawback to the procedure is that after a large 
number of iterations $J$ many of the ``ancestors'' will die out, leaving 
no descendants, which leads to a progressive loss of statistical 
accuracy. Thus it is even more crucial in this connection to use a good 
guiding wavefunction $\Psi_{\text{T}}$ so as to minimize ``branching''.   

\subsection{Timelike Correlations}
\label{subsec: tim}

In order to estimate mass gaps in this model, we have again used the 
forward-walking technique to measure correlations between operators at 
different times. The mass gaps can then be estimated from the decay 
constants for these correlation functions.

A timelike correlator $\langle Q_{1}(\tau) Q_{2} (0) \rangle_{0}$ can be 
found from the following equations:
\begin{equation}
<Q_{1}(\tau)Q_{2}(0) >_{0} = \frac{< \Phi_{0} |Q_{1}e^{-H\tau}Q_{2}| 
\Phi_{0}>}{ < 
\Phi_{0}| \Phi_{0} >} 
\label{eq17}
\end{equation}
\begin{equation}
\stackrel{\textstyle \sim}{\scriptscriptstyle J \rightarrow \infty}
 \frac{< \Psi_{\text{T}} |K^{J}Q_{1}K^{N}Q_{2}| \Phi_{0}>}{ < 
\Psi_{\text{T}}|K^{J+N}| \Phi_{0} >}
\label{eq18}
\end{equation}
\begin{equation}
= \frac{ \sum \tilde{K}^{J}
Q_{1}\tilde{K}^{N}Q_{2}\tilde{\Phi}_{0}}{ \sum \tilde{K}^{J+N}
  \tilde{\Phi}_{0}}
\label{eq19}
\end{equation}
where $N\Delta\tau = \tau$. This can be implemented in much the same
way as an expectation value (assuming both $Q_{1}$ and $Q_{2}$ are
diagonal in the weak-coupling representation). At the beginning of the
measurement, record the `ancestor' configurations as in Sec.\ 
(\ref{subsec: for}). Then allow the ensemble to propagate for time 
$\tau$, {\it with the branching process turned off} so that each state 
retains its identity. At the end of time $\tau$, record the initial 
value $Q_{1}$ and the final value $Q_{2}$. Propagate each state for a 
further $J$ iterations as before, and then average, weighting each 
`ancestor' state according to the forward-walking prescription of Sec.\ 
(\ref{subsec: for}).

\section{Results}

Simulations were carried out for $L \times L$ lattices up to $L = 16$ 
sites, using runs of 10,000 walkers over 50,000 iterations. At each
iteration several sweeps of the lattice were performed, after which 
Runge smoothing (the branching (combining) of high (low) weight walkers) 
was imposed on the walker population. The average percentage of walkers 
branched/combined per iteration depends on a number of factors: clearly, 
the time step $\Delta \tau$ and number of lattice sweeps performed per 
iteration can be adjusted in order to control the extent of the 
branching. We wish to choose values of $\Delta \tau$ which are 
sufficiently small that time discretization errors can be made 
negligible. Furthermore, the number of sweeps performed per iteration 
was chosen so that essentially uncorrelated measurements could be made 
roughly every 150 iterations. We found that, for the lattice sizes $L$ 
and couplings $x$ considered, this could be achieved in such a way that 
on average no more than 10\% of the walkers were branched/combined per 
iteration. However, the extent of the smoothing per iteration depends on 
the quality of the variational guiding function. Generally, the
two-parameter guiding function (\ref{eq13}) works better for 
smaller couplings $x$ and smaller lattice sizes $L$. Under the most 
strained conditions considered ($x = 4$, $L = 16$) around 10\% of the 
walkers were branched/combined per iteration for the values of 
$\Delta \tau$ chosen and the number of sweeps per iteration required to 
get independent measurements, whereas for $x \leq 1$ the ratio was a 
fraction of a percent. The first 2000 iterations were discarded to allow 
for equilibration. Two values of the time step $\Delta \tau$ were used, 
differing by a factor of 5 (e.g.\ $\Delta\tau = 0.02$ and 
$\Delta\tau = 0.005$). The number of sweeps performed for the smaller 
$\Delta \tau$ value was 5 times as large as that for the larger value 
and as a result measurements were equally uncorrelated in the two cases. 
The typical number of sweeps used per iteration, e.g.\ for $\Delta \tau 
= 0.02$ and $\Delta\tau = 0.005$ was 1 and 5 respectively. Linear 
extrapolations were made to the limit $\Delta\tau = 0$. The linear 
dependence on $\Delta \tau$ has been checked previously \cite{hamer1}.

\subsection{Ground-state Energy}

\begin{figure}[htbp]
%%%
 \centerline{\epsfxsize=8.4cm \epsfbox{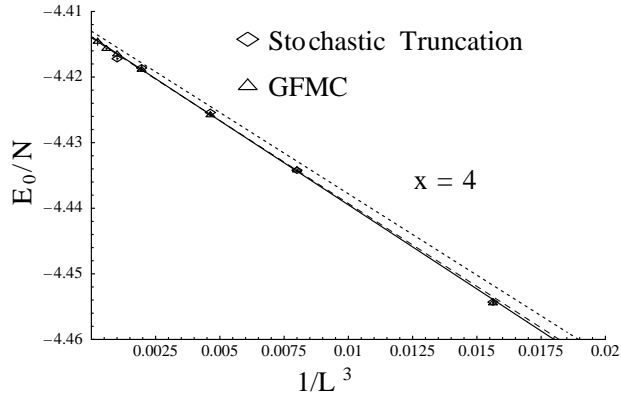}}
\caption{
Ground-state energy per site graphed against $1/L^{3}$, where $L$ is the
lattice size, at $x = 4$. The triangles show our present estimates 
(GFMC), the diamonds are from the stochastic truncation 
calculations from ref.\ \protect\cite{hamer1}. The solid and dashed 
lines are least squares fits to the GFMC and stochastic truncation 
results respectively. The dotted line is from the predictions of 
weak coupling theory (see text).
}
\label{fig2}
\end{figure}

Results for the ground-state energy per site $E_0 / L^2$ at coupling
$x = 4$ are graphed against $1/L^{3}$ in Fig.\ \ref{fig2}. They agree 
within errors with those of the earlier study \cite{hamer1} up to
$L = 10$, and extend them to $L = 16$. It can be seen that the data
are fitted quite well by the form
\begin{equation}
E_0 / L^2 = -4.414(1) -\frac{2.53}{L^{3}},
\end{equation}
which compares very well with effective Lagrangian theory and the
weak-coupling series prediction \cite{hamer4,hamer1}
\begin{equation}
E_0 / L^2 = -4.413(2) -\frac{2.48(1)}{L^{3}},
\end{equation}
The estimated bulk value $-4.414(1)$ compares well with estimates of
$-4.43(2)$ from strong-coupling series \cite{series}, $-4.415(6)$ 
from a $t$-expansion \cite{hamer2}, and $-4.412$ from the coupled
cluster method (CCM) \cite{bishop}.

\subsection{Wilson Loops}

\begin{figure}[htbp]
%%%
 \centerline{\epsfxsize=8.4cm \epsfbox{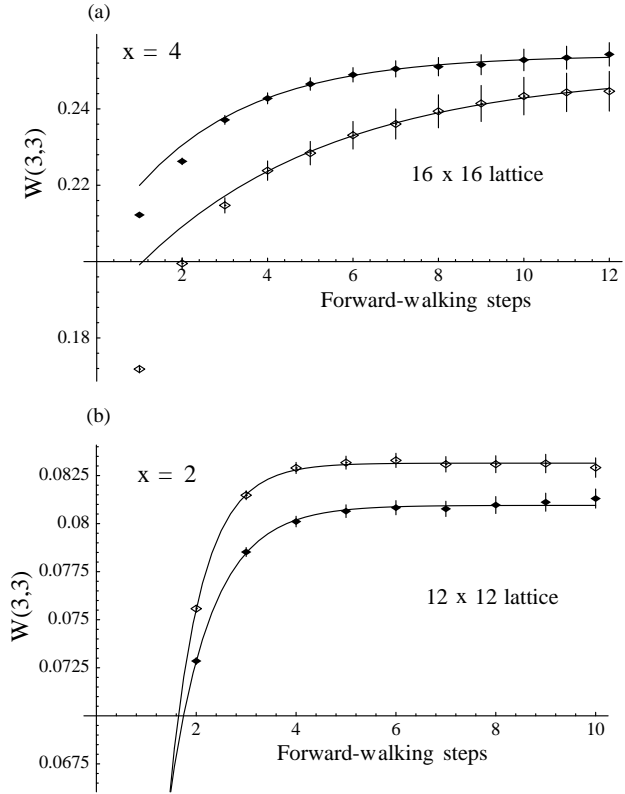}}
\caption{
(a) Forward-walking weighted averages of the Wilson loop $W(3,3)$ at 
coupling $x = 4$ and lattice size $L = 16$, as a function of the number 
of forward-walking steps. Each forward walking step involves 20
MC iterations which in turn consist of 3 lattice sweeps, with a time
step of $\Delta \tau = 0.005$. Open diamonds are results obtained from a
1-parameter guiding function, filled diamonds from a 2-parameter guiding
function. (b) Convergence of forward-walking weighted averages for the 
$3 \times 3$ Wilson loop for $x = 2$ on the $12 \times 12$ lattice. Each
forward walking step consists of 16 MC iterations. The open diamonds
are for $\Delta \tau = 0.05$ (1 lattice sweep per MC iteration) and
the closed diamonds are for $\Delta \tau = 0.01$ (5 lattice sweeps per
MC iteration).
}
\label{fig3}
\end{figure}

The expectation values of Wilson loops were computed using the
forward-walking method outlined in section \ref{subsec: for}. 
Measurements of the observables were made in cycles starting around 
every 150 iterations, with typically 10 forward-walking weighted 
averages over the ensemble taken at time steps of 12--15 iterations.
For each time 
step the weighted averages were then block averaged over successive 
measurement cycles. Fig.\ \ref{fig3}(a) shows the forward-walking 
convergence of the $3 \times 3$ Wilson loop $W(3,3)$ measured on the
$16 \times 16$ lattice at coupling $x = 4$ for a trial run involving 
1200 walkers, 20,000 iterations, a time step $\Delta \tau = 0.005$, 3 
sweeps of the lattice per iteration, with 12 forward-walking weighted 
averages being taken every 20 iterations in measurement cycles made 250 
iterations apart. Results are shown for two different guiding functions, 
using the 1-parameter and 2-parameter forms respectively. It can be seen 
that in both cases the data relax exponentially towards a common 
equilibrium value, which can be estimated by making an exponential fit 
to the data. The equilibrium value is then taken as the final result for 
the Wilson loop. It can also be seen that the 2-parameter form reduces 
the variance and produces much more rapid convergence to the asymptotic 
value.

Typical results from a production run (using the 2-parameter guiding 
function) are shown in Fig.\ \ref{fig3}(b). In this case $L = 12$,
$x = 2$ and 50,000 iterations are performed with 10,000 walkers. 
Measurement cycles involving 10 forward-walking weighted averages 16 
iterations apart were started every 165 iterations. Results for two 
values of $\Delta \tau$ (0.05 and 0.01), using 1 and 5 lattice sweeps 
per iterations respectively, are shown.

\begin{figure}[htbp]
%%%
 \centerline{\epsfxsize=8.4cm \epsfbox{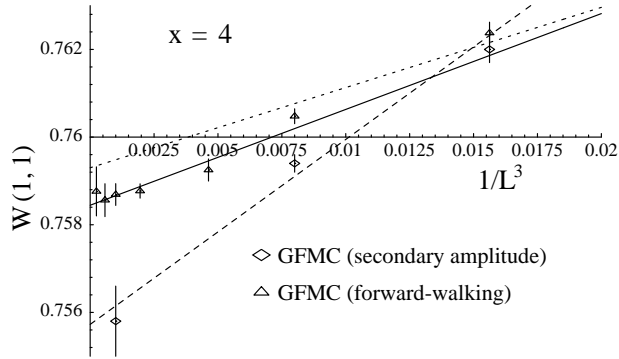}}
\caption{
The mean plaquette value $W(1,1)$ as a function of $1/L^{3}$, at
$x = 4$. The triangles are our present forward-walking GFMC results;
the diamonds are GFMC results from ref.\ \protect\cite{hamer1}
using the ''secondary amplitude'' method of evaluating expectation
values. The solid and dashed lines are least squares fits to the
forward-walking and secondary amplitude results respectively.
The dotted line is from the predictions of weak coupling theory
(see text).
}
\label{fig4}
\end{figure}

\begin{figure}[htbp]
%%%
 \centerline{\epsfxsize=8.4cm \epsfbox{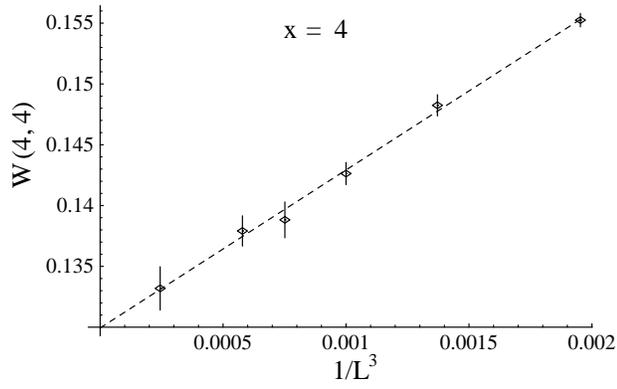}}
\caption{
Estimates of the $4 \times 4$ Wilson loop $W(4,4)$ at $x = 4$, as a 
function of $1/L^{3}$. The dashed line is a least squares linear fit
to the data.
}
\label{fig5}
\end{figure}

Fig.\ \ref{fig4} displays the values for the `mean plaquette'
$P = W(1,1)$ at coupling $x = 4$, graphed against $1/L^{3}$. There is 
evidently a discrepancy here between our present results and those of 
\cite{hamer1}, obtained using the secondary amplitude technique.
Either the errors in \cite{hamer1} have been underestimated for the 
larger lattices and couplings, or the discrepancy could have arisen
from bias in the secondary amplitude technique \cite{hamer3,samaras}.
Our present results show a consistent finite-size scaling behaviour
for $P(L)$:
\begin{equation}
P(L) = 0.7584(4) +\frac{0.22}{L^{3}},
\end{equation}
to be compared with the weak-coupling series prediction \cite{hamer4}
\begin{eqnarray}
P(L) & = & -\frac{1}{2}\frac{d}{dx}[ E_0 / L^2] \\
     & = & 0.7593(3) +\frac{0.183(3)}{L^{3}} , 
\end{eqnarray}
obtained using the Hellmann-Feynman theorem. The agreement is once again 
quite good. Extrapolation to the axis gives a bulk limit of
$P = 0.7584(4)$, to be compared with strong-coupling series
\cite{series} 0.80(3), the $t$-expansion \cite{morningstar} 0.757(4),
and the CCM \cite{bishop} 0.7585.
The results for other Wilson loops scale similarly with lattice size.
For example, the finite-size scaling of the $4 \times 4$ Wilson loop is 
shown in Fig.\ \ref{fig5}.

\begin{figure}[htbp]
%%%
 \centerline{\epsfxsize=8.4cm \epsfbox{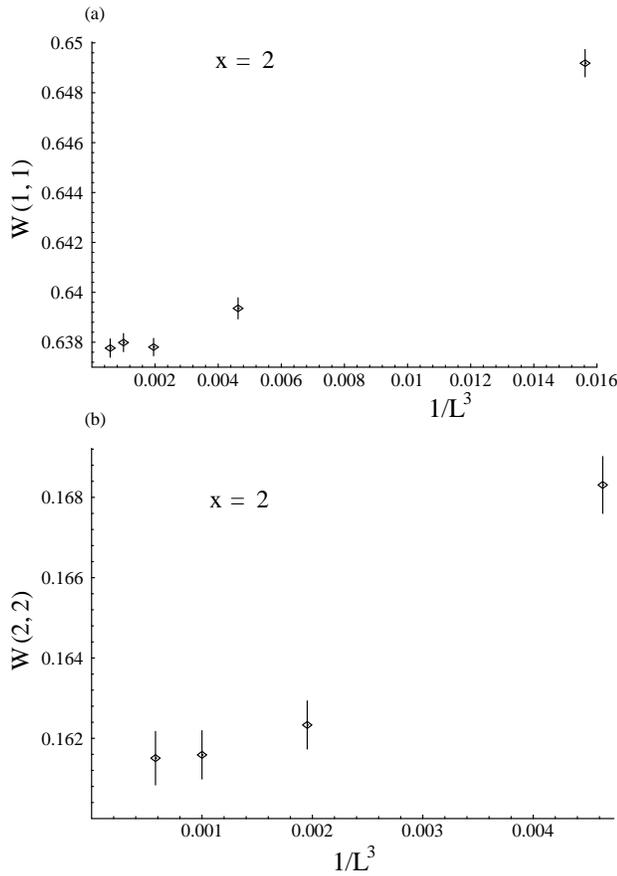}}
\caption{
(a) The mean plaquette, $W(1,1)$ and (b) the $2 \times 2$ Wilson loop as 
functions of $1 / L^3$ for the $x = 2$ case.
}
\label{fig6}
\end{figure}

Of course, because the model has a gap for any finite value of $x$, one 
should expect to see a crossover from this algebraic (free-photon 
theory) scaling to exponential convergence for large enough $L$. Indeed, 
for smaller values of the coupling, exponential scaling sets in rapidly 
enough that essentially bulk results can be obtained on relatively small 
lattices. This is illustrated in Fig.\ \ref{fig6} in the case $x = 2$ 
where we see a definite crossover from algebraic to exponential scaling 
for the mean plaquette and the $2 \times 2$ Wilson loop when $L$ reaches 
around 10.

Table \ref{table1} lists the final estimates we have obtained for the 
bulk ground-state energy per site and Wilson loop values at some 
selected couplings.

\subsection{String Tension}

\begin{figure}[htbp]
%%%
 \centerline{\epsfxsize=8.4cm \epsfbox{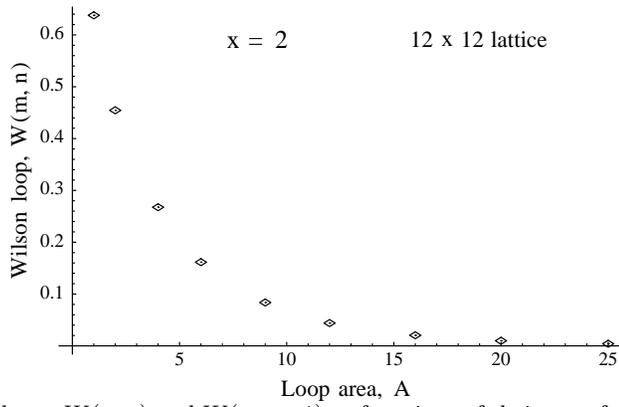}}
\caption{
Wilson loops $W(n,n)$ and $W(n,n-1)$ as functions of their area for
$x = 2$ on the $12 \times 12$ lattice.
}
\label{fig7}
\end{figure}

\begin{figure}[htbp]
%%%
 \centerline{\epsfxsize=8.4cm \epsfbox{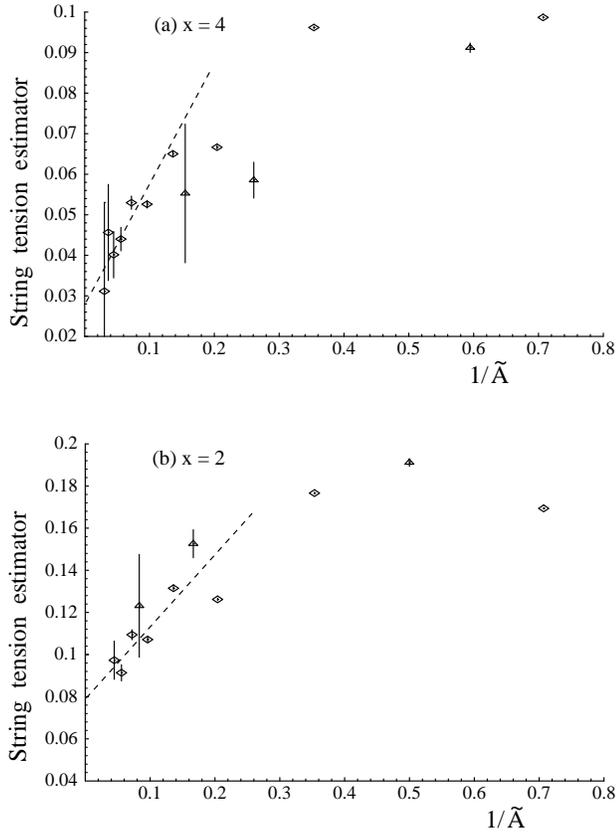}}
\caption{
`Two-point' string tension estimates $R_{n'}$ (diamonds) and Creutz 
ratios $R_{n}$ (triangles) for (a) $x = 4$ (using extrapolated
$L = \infty$ estimates of the Wilson loops) and (b) $x = 2$ (using
$L = 12$ estimates of the Wilson loops), graphed
against $1 / \tilde{A}$, where $\tilde{A}$ is the average area of 
the Wilson loops used to form the ratios.
}
\label{fig8}
\end{figure}

In a confining model, the Wilson loops are expected to behave
asymptotically as
\begin{equation}
W(m,n) \sim \exp{[-KA]}
\end{equation}
where $A = mn$ is the area of the loop and $K$ is the string tension. In 
Fig.\ \ref{fig7} we illustrate this behaviour for the $x = 2$ case on 
the $12 \times 12$ lattice by plotting the Wilson loops $W(n,n)$ and 
$W(n,n-1)$ as a function of area. The standard estimator of the string 
tension is the Creutz ratio
\begin{equation}
K \sim R_n = - \ln \left[
\frac{ W(n,n) W(n-1,n-1) }{ W(n,n-1)^{2} }
\right]
\end{equation}
Fig.\ \ref{fig8}(a) shows the Creutz ratios $R_n$ graphed against
$1 / \tilde{A}$ (where $\tilde{A} \equiv n (n - 1)$ is the 
average area of the loops used to form the ratio) at $x = 4$, from which 
we obtain the rough estimate
\begin{equation}
K = 0.05(2)
\end{equation}

Because the Creutz ratios decrease very rapidly with $n$, and the 
relative error correspondingly increases, it is very difficult to 
determine a reliable value for the string tension. Given the limited 
number of data points, it is probably more fruitful to use a
two-point estimator
\begin{equation}
K \sim R_{n}' = - \frac{1}{n}
\ln \left[ \frac{ W(n,n) }{ W(n,n-1) } \right],
\end{equation}
formed from successive pairs of Wilson loops. Included in Fig.\ 
\ref{fig8}(a) are plots of these estimators against 
$1/\tilde{A}$. Unfortunately, the two-point estimators display 
oscillations (which the Creutz ratios were designed to eliminate) so it 
is again difficult to perform a linear extrapolation to obtain an 
estimate of the string tension. Nevertheless, a linear fit through the 
points gives an estimate of the string tension of $K \approx 0.03$,
somewhat lower than the result above.

We have performed a similar analysis for $x = 2$, working with the
$12 \times 12$ lattice which is sufficiently close to the bulk limit. 
The results for the string tension estimators are shown in Fig.\ 
\ref{fig8}(b). In this case a linear extrapolation of the two-point 
estimators yields an estimate of $K \approx 0.08$ for the string tension. 

The string tension has previously been calculated as an energy per unit
length, $\sigma$. The best available values at $x=2$ are $\sigma = 0.28(1)$
 from stochastic truncation \cite{hamer1}, and $\sigma = 0.282(2)$ from an 
`exact linked cluster expansion' \cite{irving1}. 
The quantities are related by
\begin{equation}
\sigma = v K 
\end{equation}
where $v$ is the speed of light, estimated as $v = 2.27$  at
$x = 2$ from the weak-coupling expansion \cite{hamer4}. Hence
$\sigma = 0.28(1)$ corresponds to $K = 0.12(1)$. 
This is in rough agreement with the result above.

We also attempted the analysis for smaller values of $x$. However,
because the Wilson loops decay extremely rapidly in these cases,
it is not possible to obtain enough data points to make sensible
extrapolations.  

\subsection{Mass Gaps}

\begin{figure}[htbp]
%%%
 \centerline{\epsfxsize=8.4cm \epsfbox{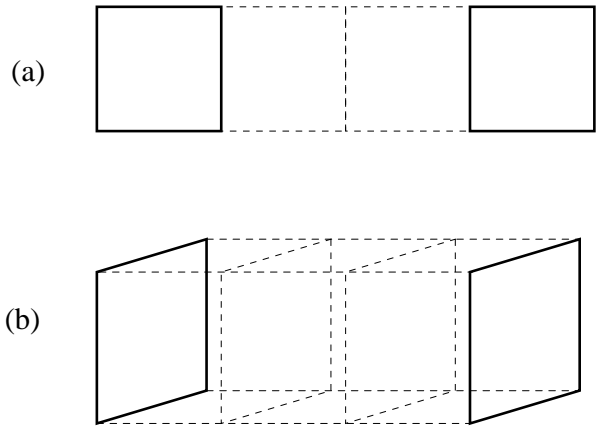}}
\caption{
(a) Two plaquettes ``edge on'' on a lattice with two spatial dimensions. 
(b) ``Face on'' plaquettes on a lattice with three spatial dimensions.
}
\label{fig9}
\end{figure}

We have attempted to estimate mass gaps from the exponential decay of
correlation functions, as is done in the Euclidean approach. The
lowest-lying excitations in the strong-coupling limit are the
single-plaquette excitations, antisymmetric or symmetric under
reflections respectively. At first we attempted to use the spatial
correlations between plaquette operators for this purpose; but the
signal from the connected spatial correlations turns out to be very
small, and can even change sign, giving no good exponential decay
signal. We can understand this problem as follows:
: we are restricted to a plane in the spatial directions, and can
therefore only measure correlations between plaquette operators ``edge
on'' to each other, as illustrated in Fig.\ \ref{fig9}(a), as opposed to 
the ``face on'' configuration in Fig.\ \ref{fig9}(b). This means there 
is very poor overlap with the configurations of interest in the 
intermediate state.

We are thus compelled to fall back on the correlations in imaginary time
$\tau$, which do correspond to ``face on'' plaquettes as in Fig.\ 
\ref{fig9}(b). Using the method of \ref{subsec: tim}, we have measured 
correlation functions
\begin{equation}
f(\tau) = \langle(\sum_{P}P(0))(\sum_{P'}P'(\tau))\rangle_{0}
\end{equation}
where the plaquette operators have been summed over all positions to
project out the zero-momentum intermediate states, and the sums are
either symmetric or antisymmetric under reflections.

As with the Wilson loop calculations, typically 50,000 iterations were 
performed with 10,000 walkers and forward-walking measurment cycles 
were started every 150 iterations. As mentioned, at the start of each 
measurement cycle, Runge smoothing was switched off so that no walkers 
were created or destroyed while timelike correlation functions were 
calculated. Between 10 and 20 timelike correlations (between the 
plaquette operators at the start of the measurement cycle $\tau = 0$ and 
later times $\tau$)were measured  at intervals of one or two iterations.
 For each of 
the timelike measurements the ``physical time'' $\tau$ is given by:
$$
\tau = \Delta \tau \times N_{\text{SWEEP}} \times
N_{\text{INTERVAL}}
$$
where $\Delta \tau$ is the basic timestep, $N_{\text{SWEEP}}$ is the 
number of sweeps of the lattice and $N_{\text{INTERVAL}}$ denotes the 
number of iterations between the calculation of initial and final 
plaquette operators.

At this point Runge smoothing was switched on again and forward-walking 
weighted averages of the timelike correlations were calculated over most 
of the measurement cycle at intervals of 10--20 iterations. Again the 
weighted averages were block averaged over all measurement cycles, and 
exponential fitting was used to find the forward-walking relaxed 
estimates of the timelike correlation functions. As with the Wilson 
loops, two values of the time step $\Delta \tau$ were used and linear 
extrapolation was used to obtain $\Delta \tau = 0$ estimates of the 
correlators. Again, the two values of $\Delta \tau$ used differed by a 
factor of 5, and 5 times as many lattice sweeps were performed per 
iteration for the smaller $\Delta \tau$ so that the ``physical'' times 
$\tau$ over which the correlations $f(\tau)$ were measured were the same 
(i.e.\ $\Delta \tau \times N_{\text{SWEEP}}$ was kept fixed). 

\begin{figure}[htbp]
%%%
 \centerline{\epsfxsize=8.4cm \epsfbox{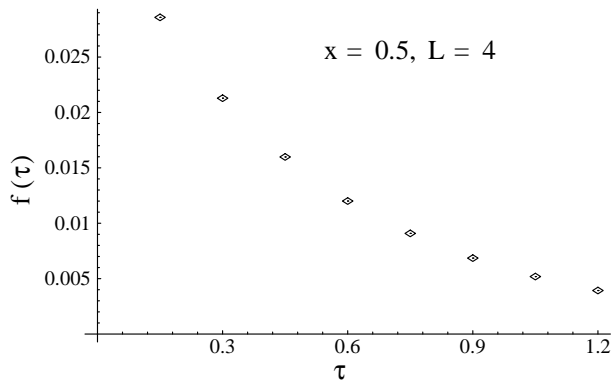}}
\caption{
Timelike correlation function $f(\tau)$ (for antisymmetric states) as a
function of imaginary time $\tau$, for coupling $x = 0.5$ and lattice 
size $L = 4$.}
\label{fig10}
\end{figure}

\begin{figure}[htbp]
%%%
 \centerline{\epsfxsize=8.4cm \epsfbox{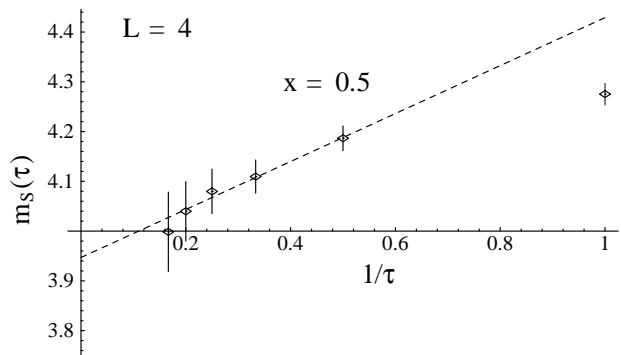}}
\caption{
``Effective mass'' $m_{\text{S}}(\tau)$ for the symmetric state as a 
function of $1 / \tau$ for coupling $x = 0.5$ and lattice size $L = 4$. 
(Here $\tau$ is measured in terms of MC iterations which correspond to a 
physical time of 0.15).
}
\label{fig11}
\end{figure}

A typical plot of the resulting estimates for the timelike correlation 
function $f(\tau)$, as a function of $\tau$,
is shown in Fig.\ \ref{fig10} for the antisymmetric 
correlator with $x = 0.5$ and $L = 4$. It can be seen that $f(\tau)$ 
shows a smooth exponential decay with $\tau$. Defining an `effective 
mass' for each pair of points by
\begin{equation}
m(\tau) = -\ln[ f(\tau) / f(\tau - 1) ],
\end{equation}
one finds that the effective mass decreases somewhat with $\tau$, as
might be expected. An {\it ad hoc} fit in $1 / \tau$, illustrated in
Fig.\ \ref{fig11}, gives a final estimate for the mass of the 
corresponding intermediate state. These values are listed as a function 
of lattice size in Table \ref{table2} for couplings $x = 0.5$, 1 and
2. For the higher couplings the exponential decay is so slow that we 
have been unable to obtain useful results for the mass. To do so would 
require measurements to be made over substantially larger time scales 
with significantly greater accuracy. Unfortunately, our results were not 
sufficiently accurate to sensibly resolve the form of the lattice size 
dependence for the couplings studied. In Table \ref{table2} we list 
results of estimates of the bulk mass from other sources. It can be seen 
that for the couplings considered, the agreement between methods is 
reasonable.

\section{Summary and Conclusions}

In this paper we have shown how the standard `forward-walking' methods
of quantum many-body theory \cite{kalos} can be used to calculate
expectation values and correlation functions in Hamiltonian lattice
gauge theory. Accurate values for the Wilson loops have been obtained,
and their finite-size scaling behaviour has been demonstrated. The
string tension can then be estimated using two-point and Creutz ratios. 
The Wilson loop values dropped away too quickly with lattice size to 
give accurate string tensions at small couplings $x$; but this problem 
could presumably be rectified by looking at timelike
correlators---either timelike Wilson loops, or timelike correlations 
between Polyakov loops.

We have also shown how mass gaps can be estimated from the exponential
decay of timelike correlators. This approach is very reminiscent of the
techniques used in the Euclidean regime: our freedom of choice of the
timelike measurement interval is akin to the choice of an anisotropic
lattice in the `timelike' direction in the Euclidean case. Reasonable
results have been obtained for the masses at small couplings $x$, but 
not at large $x$.  

The accuracy of the final results for the string tension and mass gaps is
not remarkable; but it must be recalled that this is the first time that
unbiased estimates have been obtained for these quantities using a
`weak-coupling' representation and GFMC techniques. There are no doubt
many ways by which the results could be improved: for example, by the
use of improved guiding wavefunctions, or improved (``fuzzed'')
correlation functions, or improved Hamiltonian operators. Our aim here has been to `prove the concept', that forward-walking techniques can give reliable estimates for the correlation functions. In future work
we plan to apply similar methods to the (3+1)D $SU(3)$ Yang-Mills 
theory.

\section*{Acknowledgments} This work is supported by the Australian 
Research Council. Calculations were performed on the SGI Power Challenge 
Facility at the New South Wales Center for Parellel Computing and the 
Fujitsu VPP300 vector machine at the Australian National Universtiy 
Supercomputing Facility.

\newpage

\begin{table}
\begin{tabular}{lll}
$x$ & $c$ & $d$ \\
\hline
0.5 & 0.239 & 0.0088 \\
1   & 0.409 & 0.033  \\
2   & 0.585 & 0.089  \\
4   & 0.771 & 0.150  \\
\end{tabular}
\caption{
Optimal values of the coefficients $c$ and $d$ used in the guiding 
function (\protect\ref{eq13}).
}
\label{table0}
\end{table}

\begin{table}
\begin{center}
$x = 1$
\end{center}
\smallskip
\begin{tabular}{llll}
& $L = 4$   & $L = 6$   & $L = 8$  \\
\hline
$E_0 / L^2$ & $-0.46520(3)$ & $-0.465127(2)$ & $-0.465065(2)$ \\
$W(1,1)$ & 0.4337(1) & 0.43350(7) & 0.43346(6) \\
$W(1,2)$ & 0.2161(1) & 0.21526(7) & 0.21521(6) \\
$W(2,2)$ & 0.0649(1) & 0.06377(7) & 0.06369(6) \\
$W(2,3)$ &    ---    & 0.01946(5) & 0.01927(4) \\
$W(3,3)$ &    ---    & 0.00391(5) & 0.00378(5) \\
$W(3,4)$ &    ---    &    ---     & 0.00077(3) \\
\end{tabular}
\begin{center}
$x = 2$
\end{center}
\smallskip
\begin{tabular}{llllll}
& $L = 4$   & $L = 6$   & $L = 8$   & $L = 10$  & $L = 12$  \\
\hline
$E_0 / L^2$ & $-1.58249(9)$ & $-1.5707(6)$ & $-1.56993(5)$ &
$-1.56996(3)$ & $-1.56998(3)$ \\
$W(1,1)$ & 0.6492(3) & 0.6394(2) & 0.6378(2) & 0.6380(2) & 0.6378(2) \\
$W(1,2)$ & 0.4772(3) & 0.4572(2) & 0.4544(2) & 0.4545(2) & 0.4545(2) \\
$W(2,2)$ & 0.3134(3) & 0.2726(2) & 0.2675(2) & 0.2673(2) & 0.2675(2) \\
$W(2,3)$ &   ---     & 0.1683(2) & 0.1623(2) & 0.1616(2) & 0.1615(2) \\
$W(3,3)$ &   ---     & 0.0932(3) & 0.0855(2) & 0.0846(2) & 0.0837(2) \\
$W(3,4)$ &   ---     &   ---     & 0.0459(2) & 0.0449(2) & 0.0440(2) \\
$W(4,4)$ &   ---     &   ---     & 0.0218(2) & 0.0208(2) & 0.0205(2) \\
$W(4,5)$ &   ---     &   ---     &    ---    & 0.0097(2) & 0.0099(3) \\
$W(5,5)$ &   ---     &  ---      &   ---     & 0.0036(4) & 0.0041(5) \\
\end{tabular}
\begin{center}
$x = 4$
\end{center}
\smallskip
\begin{tabular}{llllllll}
& $L = 4$   & $L = 6$   & $L = 8$   & $L = 10$  & $L = 12$  & $L = 16$  
& $L = \infty$ \\
\hline
$E_0 / L^2$ & $-4.4545(1)$ & $-4.4258(2)$ & $-4.4189(2)$ & $-4.41653(8)$ & 
$-4.4157(1)$ & $-4.4146(6)$ & $-4.414(1)$ \\
$W(1,1)$ & 0.7624(2) & 0.7592(3) & 0.7588(2) & 0.7587(2) & 0.7586(4) & 
0.7588(6) & 0.7584(1) \\
$W(1,2)$ & 0.6338(4) & 0.6256(4) & 0.6236(2) & 0.6234(4) & 0.6223(4) & 
0.6233(7) & 0.6226(2) \\
$W(2,2)$ & 0.4980(5) & 0.4751(7) & 0.4706(4) & 0.4693(6) & 0.4660(6) & 
0.4659(9) & 0.4665(2) \\
$W(2,3)$ &   ---     & 0.3739(8) & 0.3651(4) & 0.3622(7) & 0.3575(6) & 
0.356(1)  & 0.3573(3) \\
$W(3,3)$ &   ---     & 0.2872(9) & 0.2713(5) & 0.2662(8) & 0.2598(8) & 
0.257(1)  & 0.2581(5) \\
$W(3,4)$ &   ---     &   ---     & 0.2066(5) & 0.1981(8) & 0.1925(9) & 
0.189(1)  & 0.1883(7) \\
$W(4,4)$ &   ---     &   ---     & 0.1552(6) & 0.1426(9) & 0.138(1)  & 
0.133(2)  & 0.130(1)  \\
$W(4,5)$ &   ---     &   ---     &   ---     & 0.1051(9) & 0.101(1)  & 
0.094(2)  & 0.091(1)  \\
$W(5,5)$ &   ---     &   ---     &   ---     & 0.076(1)  & 0.071(1)  & 
0.066(2)  & 0.064(2)  \\
$W(5,6)$ &   ---     &   ---     &   ---     &   ---     & 0.051(1)  & 
0.045(2)  & 0.040(3)  \\
$W(6,6)$ &   ---     &   ---     &   ---     &   ---     & 0.034(1)  & 
0.031(3)  & 0.029(5)  \\
\end{tabular}
\caption{
Estimates of the ground-state energy per site $E_0 / L^2$ and Wilson 
loop values $W(m,n)$ as functions of lattice size $L$ for various 
couplings $x$. In the $x = 4$ case we extrapolate to the bulk limit
$L = \infty$ by fitting to the form
$W_L(m,n) \sim W_\infty(m,n) + A / L^3$.
}
\label{table1}
\end{table}

\begin{table}
\begin{tabular}{lllll}
$x$ & $L$  &   $m_{\text{A}}$  & $m_{\text{S}}$ &
                                 $m_{\text{S}} / m_{\text{A}}$ \\
\hline
0.5  &   4    &   3.69(5)      &  3.95(6)    &   1.07       \\
0.5  &   6    &   3.72(4)      &  3.93(6)    &   1.06       \\
0.5  &   8    &   3.66(7)      &  4.0(1)     &   1.09       \\
0.5  & Series &   3.658375(4)  &  3.91786    &   1.07093    \\
0.5  & $t$-expansion
              &   3.66(1)      &  3.92(1)    &   1.071(1)   \\
0.5  &  CCM   &   3.668        &  3.96       &   1.080      \\ 
\hline
1.0  &   4    &   3.01(6)      &  4.05(8)    &   1.35       \\
1.0  &   6    &   3.03(6)      &  3.93(13)   &   1.30       \\
1.0  &   8    &   2.98(8)      &  3.97(11)   &   1.33       \\
1.0  &  10    &   3.06(11)     &  4.03(36)   &   1.3        \\
1.0  & Series &   2.95875(5)   &  3.755(2)   &   1.2691(5)  \\
1.0  & $t$-expansion
              &   2.96(8)      &  3.8(1)     &   1.28(2)    \\   
1.0  & CCM    &   3.01         &  4.15       &   1.38       \\
\hline
2.0  &   4    &   3.0(2)       &  5.2(3)     &   1.7        \\
2.0  &   6    &   1.8(2)       &  3.6(7)     &   2.0        \\
2.0  & Series &   1.684(5)     &  2.93(7)    &   1.76(5)    \\
2.0  & $t$-expansion
              &   1.9(4)       &  3(1)       &   2.1(2)     \\
2.0  &  CCM   &   1.78         &  5.31       &   2.98       \\
\end{tabular}
\caption{
Estimates of the symmetric and antisymmetric mass gaps $m_{\text{A}}$ 
and $m_{\text{S}}$, and the ratio $m_{\text{A}} / m_{\text{S}}$ as 
functions of the lattice size $L$ for various couplings $x$. Shown for 
comparison are the bulk estimates from strong coupling series and
$t$-expansions \protect\cite{hamer2} as well as CCM results
\protect\cite{bishop1}.
}
\label{table2}
\end{table}

\newpage

\end{document}